\documentclass[12pt]{article}
\newcommand{\qq}{$q{\overline q} \ $}
\newcommand{\uu}{$u{\overline u} \ $}
\newcommand{\dd}{$d{\overline d} \ $}

\newcommand{\KK}{$K{\overline K} \ $}
\newcommand{\st}{$s{\overline s} \ $}

\usepackage{graphicx}
\usepackage{psfrag}   
\begin{document}

\hfill\vbox{\hbox{DCPT/03/98}
            \hbox{IPPP/03/49}}
\nopagebreak

\vspace{2.0cm}
\begin{center}
\large{\bf{IN THE DEBRIS OF HADRON INTERACTIONS\\ LIES THE BEAUTY OF QCD}
\footnote{$\ $\uppercase{P}resented at the \uppercase{W}orkshop on \uppercase{G}luonic \uppercase{E}xcitations, \uppercase{T}homas \uppercase{J}efferson \uppercase{N}ational \uppercase{A}ccelerator \uppercase{L}aboratory, $.\qquad$\uppercase{N}ewport \uppercase{N}ews, \uppercase{U}.\uppercase{S}.\uppercase{A}., \uppercase{M}ay 2003}
}
\vspace{0.8cm}

\large{M.R. Pennington}

\vspace{0.6cm}

{\it Institute for Particle Physics Phenomenology,\\ University of Durham, 
Durham, DH1 3LE, U.K.\\ E-mail: m.r.pennington@durham.ac.uk}\\
     
\vspace{1.0cm}

\end{center}

\centerline{Abstract}

\vspace{0.3cm}

\small

\begin{center}
\begin{minipage}{14cm}

{Recent progress in understanding the strong physics regime
of QCD is described. The role played by condensates, particularly 
$q{\overline q}$, in breaking chiral symmetry and generating constituent  masses  for  $u$ and $d$ quarks is reviewed. The influence this has on hadrons with vacuum quantum numbers is emphasised. What we know of this sector from recent data on $\phi$-radiative decays and from $D$ decays to light hadrons is discussed. The key to further understanding is  comprehensive analyses of such data, including that planned for Hall D at Jefferson Laboratory.}
\end{minipage}
\end{center}

\normalsize

\vspace{0.5cm}
\newpage
\baselineskip=6.2mm
\parskip=2mm

\section{QCD vacuum}

To learn about the underlying theory of quark and gluon interactions we have to study hadronic collisions. In the debris of these lies the beauty of QCD. The task required to reveal this world of quarks and gluons is akin to an archaeological dig through the debris of Iraq in 2003
to unearth the civilisation of Babylon beneath. Over 30 years of digging considerable progress has been made in uncovering the strong physics aspects of QCD.

What makes QCD more interesting than QED is the nature of the vacuum. For QED the vacuum is described by perturbation theory. It is  essentially empty with a low density of particle-antiparticle pairs. In QCD because the interactions are stronger not only is the vacuum a denser sea of \qq pairs and  cloud of gluons, but so strong are the forces that condensates of quarks, antiquarks and gluons form in all colour singlet combinations. That of lowest dimension is the \qq condensate. The scale of this condensate characterises the key non-perturbative effects of light hadron systems. Being non-zero it dynamically breaks the chiral symmetry of QCD. This simultaneously ensures that the pion is the Goldstone boson of this symmetry breaking and that the corresponding scalar field is the Higgs sector of the strong interaction, responsible for the masses of all light hadrons.

The value of this condensate $\langle u{\overline u}\rangle \simeq \langle d {\overline d}\rangle$ can be found in at least three different ways~\cite{mrp-prague,mrp-krakow}, which here will be described as {\lq\lq phenomenological''}, {\lq\lq experimental''} and {\lq\lq theoretical''}. Remarkably the 3 distinct ways give consistent results. The first and oldest phenomenological method is the application of the QCD sum-rules of Shifman, Vainshtein and Zakharov~\cite{SVZ} to scalar and pseudoscalar currents. The sum-rules relate the matrix element of current correlators evaluated at low energies from hadronic data to their calculation at higher energies using the Operator Product Expansion. It is in this expansion that condensates arise.
Though such sum-rules have been studied for 25 years, recent precision has come from a better understanding of the use of contour improvement, of pinched weights in finite energy sum-rules and
technological advances in the calculation of higher order corrections in perturbative QCD (see citations in Ref.~2). Agreement between the theoretical and experimental sides of the sum-rules gives $\langle q {\overline q} \rangle \simeq - (250 \pm 25 \ {\rm MeV})^3$
at a scale of 2 GeV. Of course, such sum-rule analyses are only a consistency check on the size of condensates not an absolute determination.

Much more direct experimental confirmation is obtained by measuring low energy $\pi\pi$ scattering precisely~\cite{mrp-krakow,mrp-daphce,stern,colangelo}. Pions being the Goldstone
bosons of chiral symmetry breaking know about the size of $\langle q {\overline q} \rangle $.
This is reflected in the value of the $\pi\pi$ scattering amplitude at the symmetry point in the middle of the Mandelstam triangle $s = t = u = 4m_{\pi}^2/3$.
At this unphysical kinematic point, the size of the amplitude increases by up to a factor of 4 as the \qq condensate decreases in scale from 250 MeV~\cite{stern}. Precision measurements (to better than 10\%) of $\pi\pi$ interactions below c.m.
energies of 450 MeV have now become possible. Combining such data with dispersion relations that incorporate the important constraint of the 3-channel crossing symmetry of $\pi\pi$ scattering allows the value of the amplitude at the symmetry point to be determined and hence fixes the size of the condensate~\cite{colangelo}. While we await the precision measurement of the amplitude at threshold deduced from the lifetime of pionium~\cite{DIRAC}, one can use the difference of the phase of $S$ and $P$-wave $\pi\pi$ interactions as measured in $K\to e \nu\pi\pi$ decay, as described in detail in Ref.~4.  
Results from the BNL-E852 experiment~\cite{E852} yield a condensate of $\sim -(270\ {\rm MeV})^3$ at 2 GeV, showing that more than 90\% of the Gell-Mann-Oakes-Renner relation for $m_{\pi}^2\, f_{\pi}^2$ expanded in powers of the current quark mass is given by just the first term linear in $m_q$~\cite{leutwyler}.

For a theoretical determination  one can solve strong QCD in the continuum using the Schwinger-Dyson equations under certain plausible assumptions,
as discussed in Refs. 10,1,2. These studies enable the behaviour of gluon, ghost and quark propagators and their interactions to be investigated in the small quark mass (or chiral) limit. Here considerable progress has been made in the past decade~\cite{alkofer,rbtswilliams}. We  understand that if the effective quark-gluon coupling becomes of order unity for momenta below 500 MeV or so, chiral symmetry is broken. Then a massless current quark (or one of mass of a few MeV), that propagates almost freely over very short distances, has an effective mass of 350 MeV at distance scales of 1 fm. The strong QCD dressing of the $u$ and $d$ quark propagators turns a current quark into a constituent quark. Remarkably, this behaviour corresponds in the chiral limit to the $\langle q {\overline q} \rangle$
condensate having a scale of 250-270 MeV~\cite{maris}, in reassuring consistency with the phenomenology just discussed.

As emphasised repeatedly by Roberts and collaborators~\cite{craig}, the axial Ward identity ensures that the  \qq bound state with pseudoscalar quantum numbers is the Goldstone boson with its interactions governed by PCAC. In contrast the bound states with scalar and vector quantum numbers have masses reflecting the mass of the fully dressed (or constituent) quark. The behaviour of the gluon and ghost propagators built into these calculations can be compared with Monte Carlo lattice simulations and are in excellent agreement~\cite{alkofer2,alkofer}. While lattice calculations  can only be performed with
sizeable quark masses, the continuum Schwinger-Dyson/Bethe-Salpeter system can be computed even in the massless limit with all the essential physics of chiral logs built in. Consequently, this system provides a modelling of the chiral extrapolation~\cite{watson} so necessary to obtain physically meaningful results for light hadrons on the lattice. The successes described in Refs.~10-13 of this approach to strong physics justify the assumptions needed to truncate the Schwinger-Dyson equations and illustrate how considerable progress has been made in extending the calculability of QCD from the perturbative regime to confinement scales so crucial for light hadron phenomena.

\section{Scalar hadronic sector} 

So far we have learnt that the dynamical breakdown of the chiral symmetry of QCD generated predominantly by \uu, \dd condensates ensures that pions are  Goldstone bosons and the corresponding scalar field plays the role of the Higgs  of the strongly interacting sector, its mass and that of all light hadrons reflecting
the constituent mass of $u$ and $d$ quarks. But what is this scalar field? Is it the $f_0(400-1200)$ (or $\sigma$), or $f_0(980)$, or $f_0(1370)$, or  $f_0(1510)$, or $f_0(1720)$, or some mixture of all of these? None of these states is likely to be a pure \qq state, none likely to be pure glue, none solely $qq{\overline{qq}}$ or a \KK molecule.
All are mixtures of these, but what mixtures? This is the outstanding issue on which we try to shed a little light.

\begin{figure}[ht] 
\begin{center}
\includegraphics[width=9.5cm]{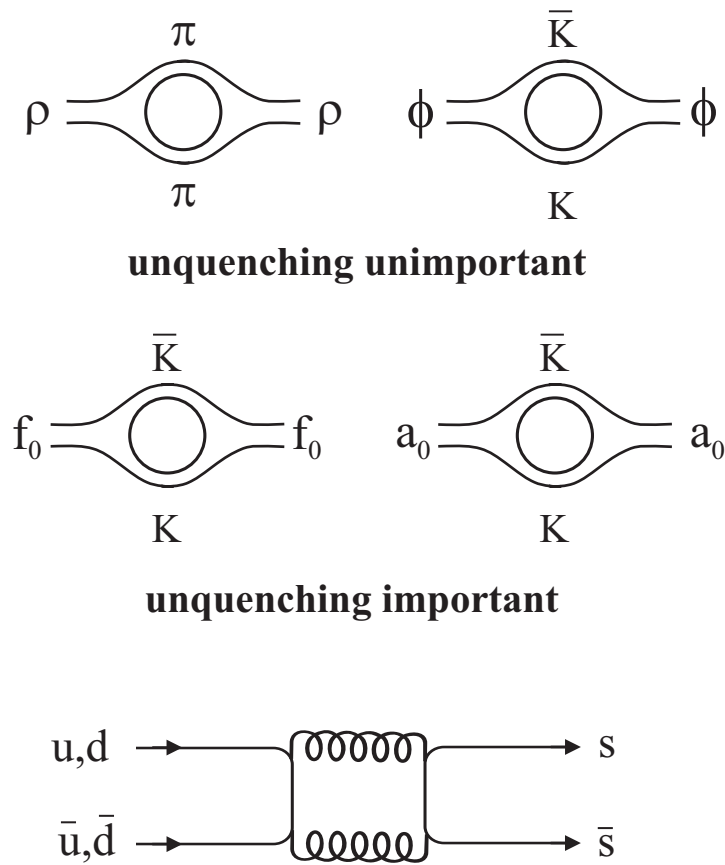}
\vspace{3mm}   
\caption{\lq\lq Unquenching'' of quark model states to make real hadrons has little effect on the vector mesons, $\rho$ and $\phi$, beyond allowing them to decay. In contrast the observed properties of the two scalar mesons $f_0(980)$ and $a_0(980)$ are produced by \lq\lq dressing''. These states then  enhance the coupling of $u{\overline u}$, $d{\overline d}$ systems to $s{\overline s}$ with 
no OZI suppression in scalar channels. \label{inter}}
\end{center}
\vspace{-3mm}
\end{figure}

The hadron states we know best, those that live the longest, are believed to be simply bound states of quarks.
States of the quark model are most easily identified with the hadrons we observe experimentally when {\it unquenching} is unimportant, Fig.~1. Thus the $\phi$ is readily seen to be an $s{\overline s}$ state and the $\rho$ and $\omega$ combinations of $u{\overline u}$ and $d{\overline d}$. This follows  from their respective decays to $K{\overline K}$, and to $2\pi$ and $3\pi$.
Though these decay modes are a crucial characteristic of their make-up, they have a relatively small effect on the states themselves. Thus only a small part of the Fock space decomposition of the physical $\phi$ is $K{\overline K}$ : it is predominantly $s{\overline s}$. This is in part because of the $P-$wave nature of its hadronic {\it dressing}. The resulting small effect  reproduces the suppression of the $1/N_c$ expansion. In contrast, $q{\overline q}$ scalar mesons are strongly disturbed by their couplings to open hadron channels~\cite{vanbeveren}, Fig.~1. Thus almost
regardless of their composition in the {\it quenched} approximation, the $f_0(980)$ and $a_0(980)$ are intimately tied to the opening of the $K{\overline K}$ threshold. Scalars change on unquenching. For them the $1/N_c$ suppression of quark loops does not occur.  The fact that these resonances, $f_0(980)$ and $a_0(980)$, couple to both $\pi\pi$/$\pi\eta$ and $K{\overline K}$, means scalar non-strange and $s{\overline s}$ states communicate, Fig.~1. The coupling of different flavour quark pairs
is not merely unsuppressed, nullifying the OZI rule in the scalar sector~\cite{geiger-isgur}, but is even enhanced. This places strange quark pairs in the vacuum under the spotlight for further study~\cite{descotes-stern}. 

 Since scalars are so intimately tied to the structure of the QCD vacuum, their nature is something we need to understand. Here we will continue to focus on the $f_0(980)$ and $a_0(980)$. It has been proposed for decades that these states have one of three possible compositions: either  a simple \qq structure
(which for the  $f_0(980)$ is
dominated by an \st component, since we know it couples strongly to \KK), or a tightly bound four quark system or a looser \KK molecule. A way to distinguish between these options is to study the $f_0$ and $a_0$ in $\phi$-radiative decays. It has been advertised by Achasov~\cite{achasov}, by Close and Isgur~\cite{close-isgur}, and by others (see citations in Refs.~19,20) that these give rise to quite distinctive branching ratios, which for $\phi\to\gamma f_0(980)$ are given in Table~1. There are analogous predictions for the $a_0(980)$: for instance, in the \KK molecule picture, where $K^+K^-$ loops are key, the ratio $BR(\phi\to\gamma a_0)/BR(\phi\to\gamma f_0)$ would clearly be one if the $a_0$ and $f_0$ were degenerate in mass.
We will see that the models used to predict the branching ratios for the 3 options shown in Table~1 are probably too simplistic, but that's for later.
\begin{table}[h]
\vspace{-1mm}
\begin{center}
\caption{\leftskip=3.1cm\rightskip=3.1cm{Predictions for the absolute rate for $\phi\to\gamma f_0(980)$ depending on the composition of the $f_0(980)$.}}
\vspace*{3mm}
\begin{tabular}{|c|c|} 
\hline
{} &{} \\[-1.ex]
  Composition     &  BR$(\phi\to\gamma f_0(980))$\\[1.5ex]
\hline
{} &{} \\[-1.ex]
 $qq{\overline{qq}}$ &$O(10^{-4})$\\[1.6ex]
 $s{\overline s}$   &$O(10^{-5})$\\[1.6ex]
 $K{\overline K}$    &$< O(10^{-5})$\\ [1.6ex]
\hline
\end{tabular} \label{tab2}
\end{center}
\vspace*{-3mm}
\end{table}

\baselineskip=6.4mm

$\phi\to\gamma\pi\pi$ and $\phi\to\gamma\pi\eta$ have recently been measured 
at $e^+e^-$-colliders: at the Novosibirsk VEPP-2M with both the SND~\cite{SND} 
and CMD-2~\cite{CMD-2} detectors and at Frascati with DAPHNE in the KLOE 
experiment~\cite{KLOEa,KLOEf}.    
 The $\pi\pi$ and $\pi\eta$ spectra, Fig.~2, show a peaking at the end of 
phase space that the experiments identify with the $f_0(980)$ and $a_0(980)$, 
respectively. 

\begin{figure}[ht]
\vspace{3mm}
\begin{center} 
\includegraphics[width=13.5cm]{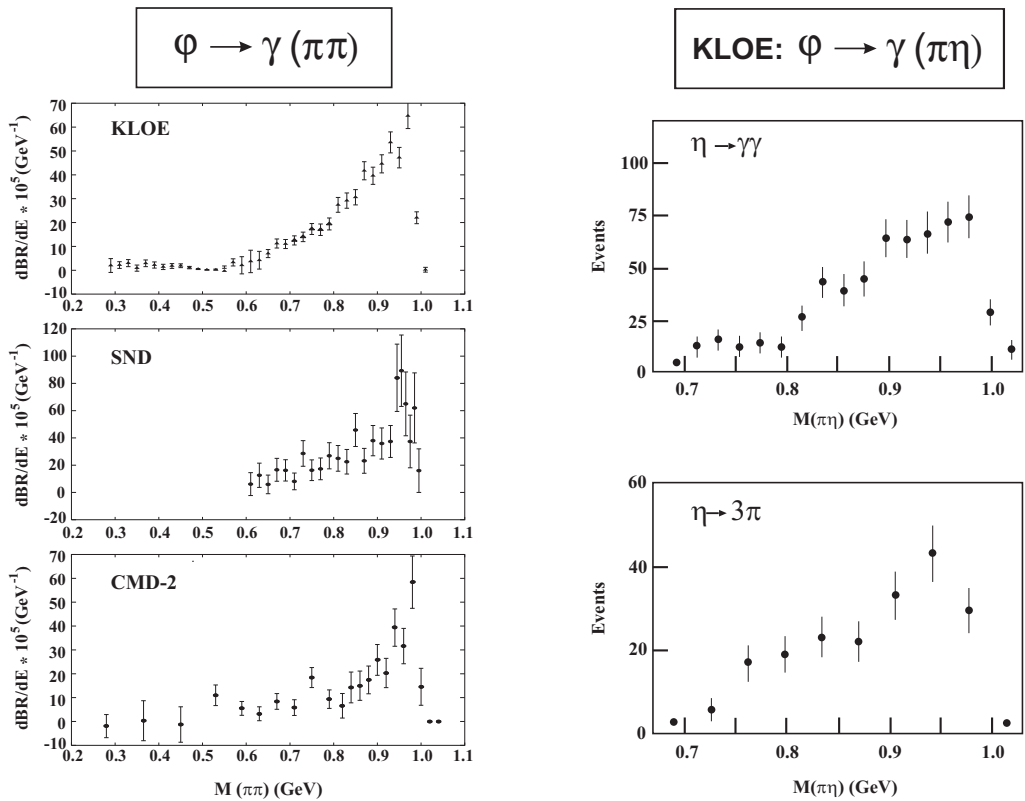} 
\caption{$\pi^0\pi^0$ and $\pi^0\eta$ mass distributions in $\phi$-radiative 
decays. The $\pi\pi$ results are from the KLOE~\protect \cite{KLOEf}, SND~\protect \cite{SND} 
and CMD-2~\protect \cite{CMD-2} detectors. The $\pi\eta$ data displayed are from 
KLOE~\protect \cite{KLOEa} in both the $\gamma\gamma$ and $3\pi$ decay modes of the $\eta$. }
\end{center}
\vspace{-3mm}
\end{figure} 

Let us look at how KLOE~\cite{KLOEf} use the data of Fig.~2 to determine the decay rate of the $f_0(980)$ to compare with the predictions of Table~1. First they fit a Breit-Wigner form to the shape of the distribution, by adjusting the mass and width of the $f_0$. Then one sees that this fails to describe the distribution at low $\pi\pi$ masses, so part of the decay system must be provided by something other than the $f_0(980)$. 
There is, of course, the background to the same $\gamma\pi\pi$ final state from the sequential decays of $\phi\to\rho\pi$ and $\rho\to\gamma\pi$. With the higher statistics from DAPHNE, the KLOE collaboration can separate out this particular decay by its distinctive angular distribution. One thus knows for the KLOE results that the $\gamma\pi\pi$ distribution (Fig.~2) is controlled by $S$-wave $\pi\pi$ interactions. As we have seen these cannot just come from $f_0(980)$ production. The remainder is taken to be the effect of the $\sigma$. The
$\sigma$'s parameters are then taken from the Fermilab E791 experiment~\cite{E791-sigma} with a mass of 478 MeV and a width of 324 MeV. The amplitudes for $\sigma$ and $f_0(980)$ production are then simply added and allowed to interfere. The resulting fit gives BR($\phi\to\gamma f_0(980)) = (4.4 \pm 0.4)\cdot 10^{-4}$  and one concludes the $f_0(980)$ is a four quark system --- see Table~1.
A~similar analysis~\cite{KLOEa} of the $\eta\pi$ channel gives a branching ratio for the $a_0(980)$ in $\phi$-radiative decay a factor of 6 smaller.
So why should we redo this analysis? 

Let us first concentrate on the isoscalar channel.
The parameters of the $f_0(980)$ are not free variables. The same resonance pole position must appear in all processes to which the state couples. Consequently, one cannot permit the mass to be reduced by 10 MeV or more, or the width of a state that is typically found to be 40-60 MeV wide cannot be allowed to be 200-250 MeV wide. These changes made by KLOE increase the branching ratio through the $f_0(980)$ by an order of magnitude. To see why, consider the key dynamics of the process.
The phase space involves a product of the $\pi\pi$ momentum and the photon momentum. But because the photon is not any massless particle, but couples through a conserved current, the decay distribution in fact involves the cube of the photon momentum, as Achasov~\cite{achasov2} has emphasised on many occasions. In terms of invariants this is proportional to $(m_{\phi}^{\,2}-M(\pi\pi)^2)^3$. With the $f_0(980)$ so close to the end of phase space, small changes in its mass dramatically alter its branching ratio. This experiment cannot determine the parameters of resonances on its own. These parameters must however be the same as those required to describe other data. Moreover, the contribution of the $\sigma$ and the $f_0(980)$ must be added in a way consistent with unitarity.
How to do this has recently been worked out by Elena Boglione and myself~\cite{mb-mrp}.

\begin{figure}[b] 
\vspace{3mm}
\begin{center}
\includegraphics[width=15.cm]{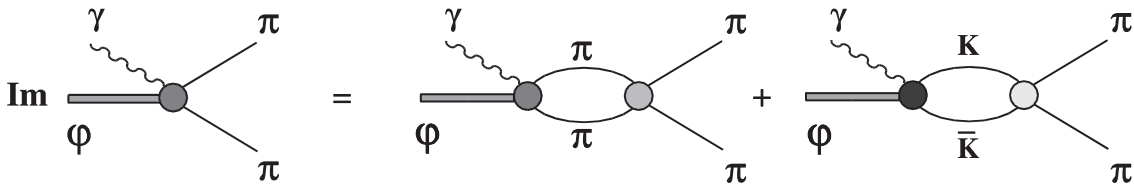} 
\vspace{3mm}  
\caption{The unitarity constraint relevant to $\pi\pi$ interactions in a definite partial wave in $\phi$-radiative decay. When there are no significant $\phi\pi$ strong interactions, or these have been removed from the data, the constraint is exact. Its solution is Eq.~(1).\label{unitarity}}
\end{center}
\end{figure} 

To proceed the basic assumption is that there are  no strong $\phi\pi$ interactions. This means that the dominant strong interaction is between the final state pions. This is the presumption implicit in any isobar modelling of a decay. With the upper  mass range fixed by the $\phi$ mass, there are very limited ways a $\pi\pi$ final state can be produced, Fig.~3. Either the $\phi$ radiates a photon leaving a $\pi\pi$ system that then interacts, or the $\phi$ radiates  a photon producing a \KK system that then interacts to produce a dipion pair. This occurs with the $\pi^0\pi^0$ system being in an isoscalar state, which is predominantly $S$-wave. 
Then coupled channel unitarity requires that the amplitude, ${F}$ for $\phi\to\gamma\pi\pi$ can be related to the basic amplitudes ${T}$ for $\pi\pi\to\pi\pi$ ($T_{11}$) and ${\overline K}K\to\pi\pi$ ($T_{21}$) by
\begin{equation}
{F}(\phi\to\gamma\pi\pi)\,=\,\alpha_1(s)\,{T}(\pi\pi\to\pi\pi)\,+\,\alpha_2(s)\,{T}({\overline K}K\to\pi\pi)\; ,
\end{equation}
\begin{figure}[t]
\begin{center}
\resizebox{1.00\textwidth}{!}{%
\begin{tabular}{cc}
\rotatebox{-90}{\includegraphics{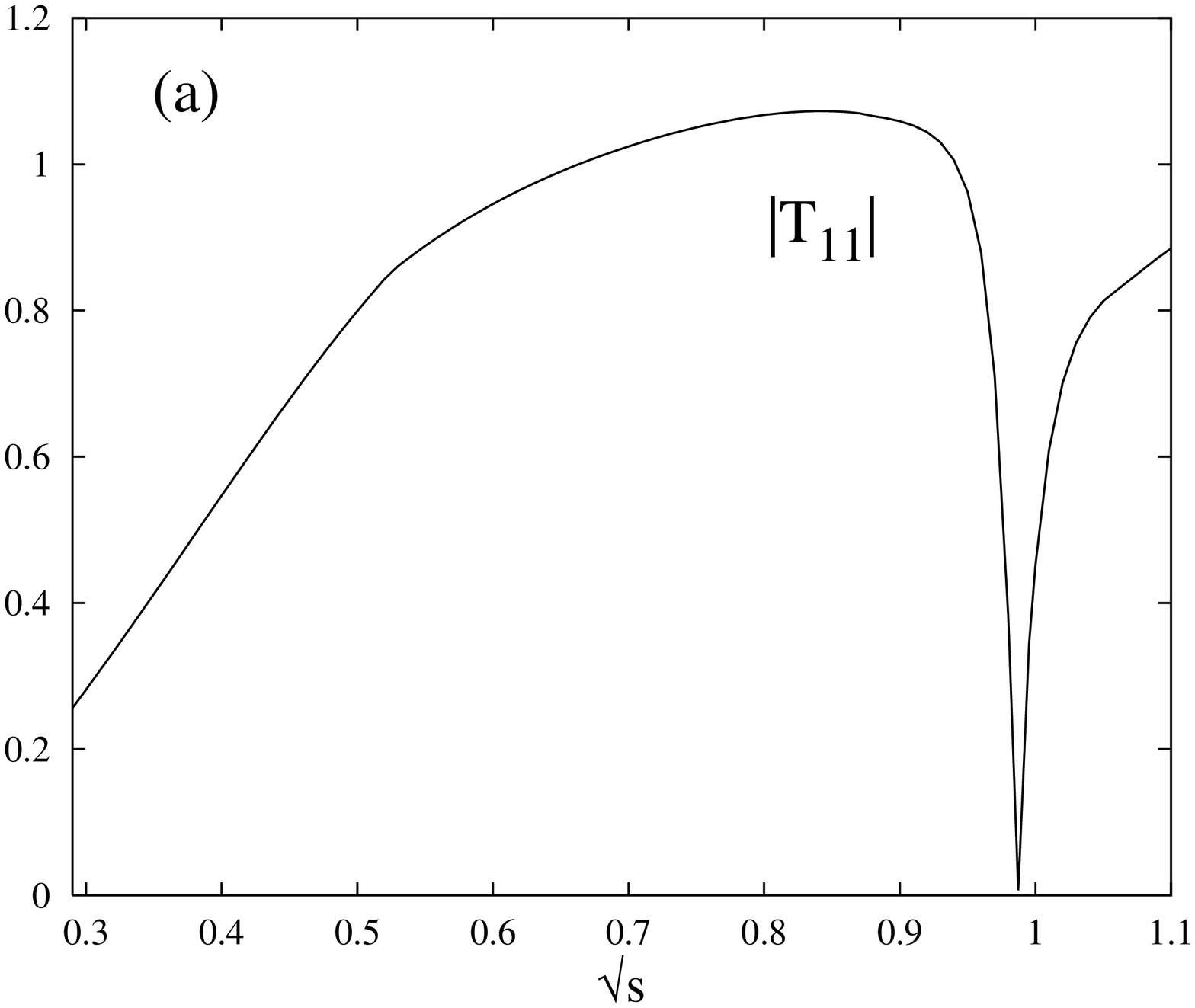}} & \rotatebox{-90}{\includegraphics{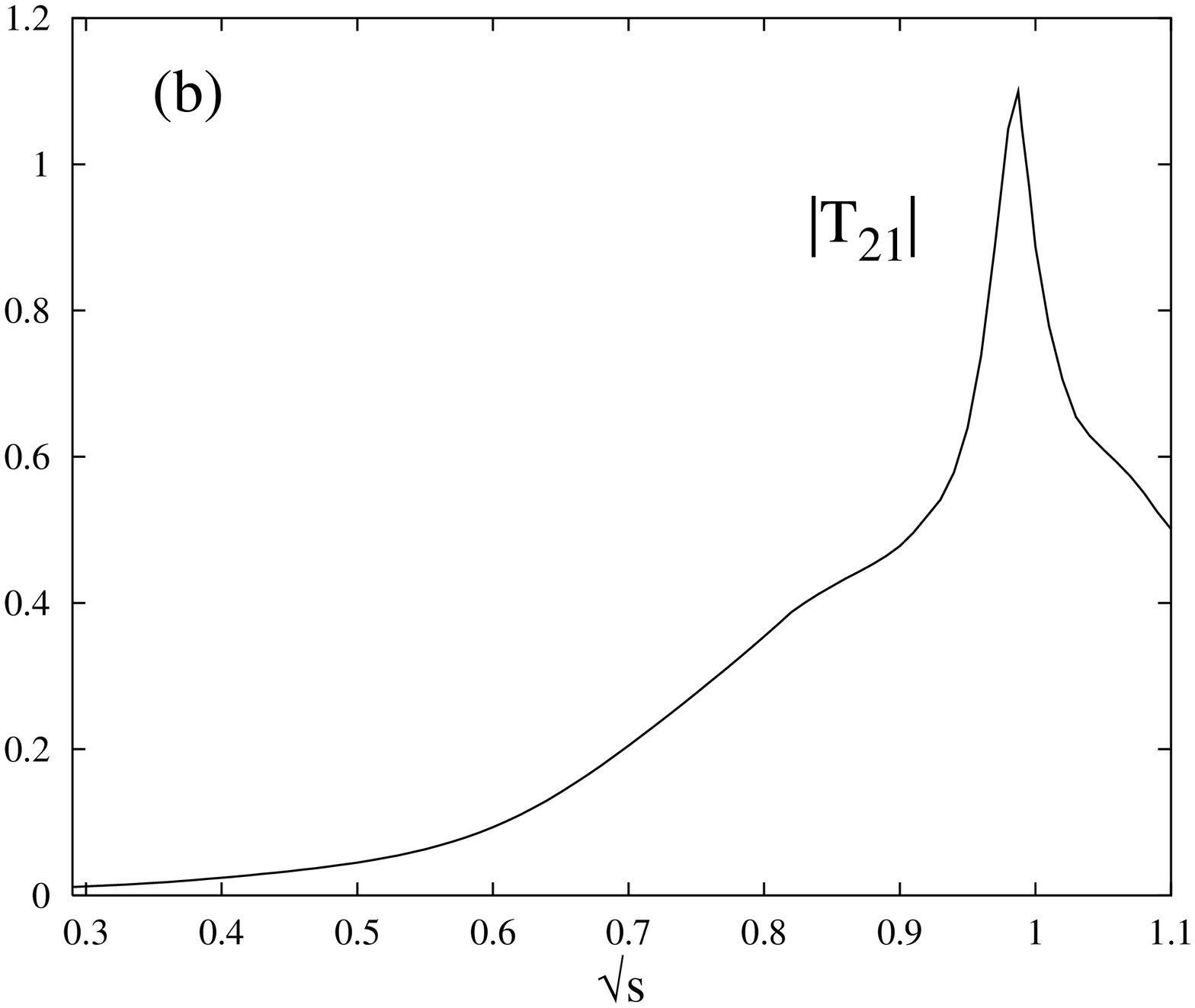}} \\
\end{tabular}
}
\caption{\label{fig-underlyingamps} 
Plots ({\bf a}) and ({\bf b}) show the $I=J=0$ ReVAMP hadronic amplitudes~\protect\cite{revamp} $|T(\pi\pi\to\pi\pi)|$ (labelled $T_{11}$) and $|T({\overline K}K\to\pi\pi)|$ (labelled $T_{21}$)
 from which $F(s)$ is constructed, according to Eq.~(1), where $\sqrt s = M(\pi\pi)$, the c.m. energy.
}
\end{center}
\end{figure} 
where $s=M(\pi\pi)^2$. With no strong $\phi\pi$ interactions, the two functions $\alpha_i(s)$ are real. They can be interpreted as the intrinsic couplings for $\phi\to\gamma\pi\pi$ and $\phi\to\gamma{\overline K}K$, respectively for $i=1,2$ (Fig.~3). Since the whole point of studying this process is because the initial state is almost 100\% \st, we expect the coupling $\alpha_2$ to be much larger than $\alpha_1$. Our analysis shows that experiment does indeed support this. However, the functions $\alpha_i(s)$ each have a factor
of $(m_{\phi}^{\, 2}-s)$, as previously explained, following from QED gauge invariance. This strongly reduces the contribution from
${T}({\overline K}K\to\pi\pi)$, which is dominated by the $f_0(980)$, and enhances the contribution from ${T}(\pi\pi\to\pi\pi)$, which is controlled by the
$f_0(400-1200)$ (or $\sigma$), and from which the $f_0(980)$ effectively decouples --- see Fig.~4.
So while there is a sizeable $f_0(980)$ component, much of the $\pi\pi$ decay distribution is produced by  $\pi\pi$ interactions outside the narrow $f_0(980)$
region.

By building into the analysis known experimental information on the scattering reactions $\pi\pi\to\pi\pi$ and $\pi\pi\to{\overline K}K$, which automatically embodies details of the $f_0(400-1200)$ and $f_0(980)$, we can use the data on $\phi\to\gamma\pi^0\pi^0$ to determine the couplings of these scalar resonances to this channel in as model-independent a way as possible. This is the purpose of the recent analysis by Elena Boglione and myself~\cite{mb-mrp}. As a simple template we 
first used the old hadronic amplitudes determined by David Morgan and~I~\cite{revamp} (called ReVAMP as explained in Ref.~27). These have the $f_0$-pole on sheet~II at $M(\pi\pi) = (988-i\cdot 23)$ MeV.
Factoring out the Adler zero and the photon momentum required by QED gauge invariance for the radiative decay process, we then have constant coupling functions, $\alpha_i(s)$, and obtain the fit shown in Fig.~5.
The quality of the fit is excellent indicating no reason to expect significant strong $\phi\pi$ interactions need be included and showing the final state $\pi\pi$ interactions in this decay are completely consistent with those from other processes built into the ReVAMP amplitudes. 
\begin{figure}[h]
\begin{center}
\resizebox{1.0\textwidth}{!}{%
\begin{tabular}{c}
\rotatebox{-90}{\includegraphics{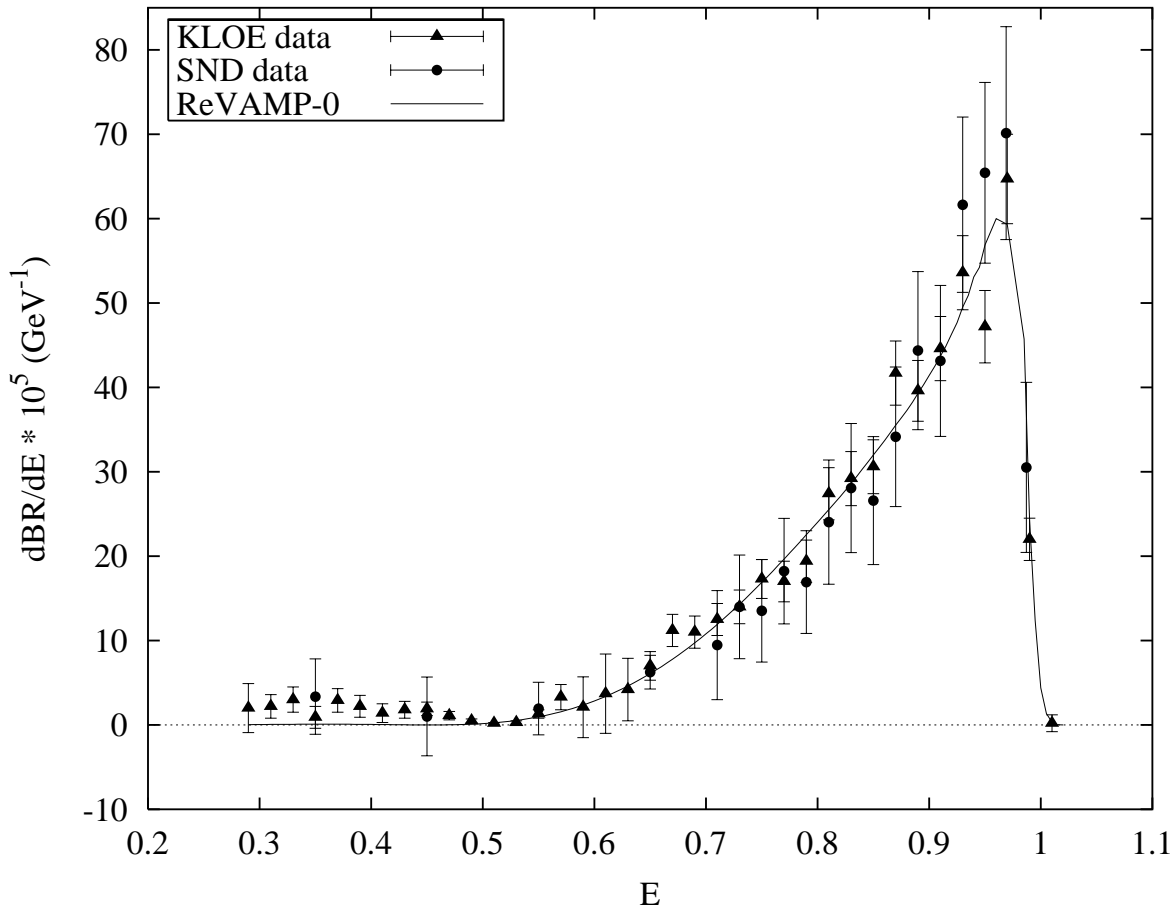}}   
\\
\end{tabular}
}
\caption{ 
Simultaneous fits to the data on the $\pi^0\pi^0$ decay distribution in 
$\phi\to\gamma\pi^0\pi^0$ from the KLOE~\protect\cite{KLOEf}  and SND~\protect\cite{SND} 
collaborations. These fits have been obtained  using the ReVAMP set of 
underlying amplitudes, 
with just 3 real parameters~\protect\cite{mb-mrp}, as described in the text. 
$E=M(\pi\pi)$.
}
\end{center}
\vspace{-3mm}
\end{figure} 

\baselineskip=6.2mm

It should be clear that the exact branching fraction for the $f_0(980)$ is exceedingly sensitive to the  position of the  corresponding pole, because of its nearness to the edge of phase space. Moreover, one does not see the $f_0(980)$ as a simple Breit-Wigner  shape (compare Fig.~4b with Fig.~5), so  in fact its branching fraction is not directly related to an experimental observable. The only well-defined quantity is the residue at the pole which truly represents the coupling $\phi\to\gamma f_0$. For a state that overlaps with another broad resonance and with a strongly coupled threshold, the idea of a branching fraction is wholly model-dependent.

The fit shown in Fig.~5 gives a $\phi\to\gamma f_0(980)$ coupling of 0.62 MeV (see Ref.~27 for the exact definition), which approximates to a branching ratio of 10\% of the total $\pi\pi$ $S$-wave decay being through the $f_0(980)$, i.e. a branching fraction of $0.31 \cdot 10^{-4}$. Looking at Table~1 we see this corresponds no longer to the $qq{\overline{qq}}$ composition inferred by KLOE~\cite{KLOEf}, but being an order of magnitude smaller is like that for an \st constitution. However, this is too simplistic. Oller~\cite{oller} has noted that the \KK molecular picture can include not just the charged kaon loops used in the predictions of Table~1, but also neutral kaons too. Then the prediction extends from $10^{-6}$ to $10^{-4}$ to encompass our result too. 

What of $\phi\to \gamma a_0(980) \to \gamma (\pi^0\eta)$? A coupled channel analysis like that for the $\pi^0\pi^0$ decay mode is not presently practicable because of lack of precision information on $\pi\eta\to\pi\eta$  and $\pi\eta\to {\overline K}K$ scattering in $S$-wave channels. However, these have none of the complication of overlapping resonances of the isoscalar $\pi\pi$ mode.
Consequently, it is more likely that such an analysis when possible will reveal a $BR(\phi\to\gamma a_0(980)$) much more similar to that found by KLOE~\cite{KLOEa}. This is the same order of magnitude as our smaller $BR(\phi\to\gamma f_0(980)$, which is something expected if there is a large \KK component of these two scalar mesons. Of course, the most simple minded interpretation of the \KK molecule picture with just charged kaon loops seeding the decay gives the ratio of branching ratios for $a_0$ and $f_0$ to be one, only if these states are degenerate in mass. Just a 10 MeV difference in their masses  changes this to 0.4 or 2, depending on the sign of the mass difference. Consequently, model predictions and analyses of data are extremely sensitive to  the $f_0(980)$ and $a_0(980)$ pole positions.

Elena Boglione and I~\cite{mb-mrp} have illustrated this by considering in our analysis of the $\pi^0\pi^0$ data the more recently determined hadronic amplitudes of Anisovich and Sarantsev~\cite{AS}. These give the $f_0$-pole 
at $M(\pi\pi)=(1024-i\cdot 43)$ MeV. Fits displayed in Ref.~27  show that this much greater width for the $f_0(980)$ results in a coupling of 1.9 MeV, which translates into a modelled branching ratio of $1.9 \cdot 10^{-4}$ ---  a factor 6 larger than using the ReVAMP amplitudes with a much narrower $f_0$ described earlier.
However, the use of Anisovich and Sarantsev amplitudes (AS) gives fits of much poorer quality. If it were not for the fact that these authors treat a far greater range of more recent data, such as that from Crystal Barrel and from GAMS (see Ref.~30 for details), than in the older ReVAMP analysis~\cite{revamp}, the quality of fit would dismiss such a  wide $f_0(980)$ as unlikely.
Nevertheless, the fact that the $\phi\to\gamma f_0$ coupling is so sensitive to the details of the $f_0$-pole position means that we need to tighten up our determination of the underlying hadronic amplitudes in this crucial mass range to be certain  of the results. This would not be the case if this same final state were to be studied in the decay of the recurrence of the $\phi$ at 1680 MeV.
This may indeed become possible in photoproduction in Hall D at Jefferson Lab.

States in the spectrum are identified as poles of the $S$-matrix. 
Crucially, their position in the complex energy plane is independent of the process in which they appear. 
Experiment does not, of course, deliver direct information about such complex poles, but only about real quantities on the real energy axis. It has been argued
\cite{michael} that this means poles are irrelevant. Only experimental observables matter and these will eventually be calculable from QCD on the lattice. Such calculations may indeed reproduce experiment precisely. However, agreement does not necessarily bring understanding. Monte Carlo lattice simulations are a {\it black box}. A perfect prediction for the $e^+e^-\to\pi^+\pi^-$ cross-section and the $J=1$ $\pi\pi$ phase-shift below 1 GeV gives no comprehension that the $\rho$-resonance with a universal pole position describes both datasets. More starkly, only the universality of poles tells us that the appearance of the dip near 1 GeV in $I=J=0$ $\pi\pi$ scattering (Fig.~4a) and the peak in $\phi\to\gamma(\pi\pi)$ (Fig.~2) are manifestations of the same $f_0(980)$. However accurately lattice QCD computations  reproduce data, only continuation into the complex energy plane brings understanding of the spectrum of hadrons and their couplings. 
Consequently, precision knowledge of the the pole positions and residues of the $f_0/a_0(980)$ is essential if we are to draw conclusions about  the composition of these states
in as model-independent way as possible --- independently
 of the modelling of $\phi$-radiative  decay  by Achasov~\cite{achasov} 
or the generation of  these states by unitarisations of chiral dynamics~\cite{oset,oller} --- just using experiment.

 A rich source of additional information on light hadron final states is being provided by the decays of heavy flavour mesons.
 That final state interactions and the way they universally appear shapes such decays has recently been studied in $D\to 3\pi$~\cite{FOCUS}. Data are typically analysed in an isobar picture~\cite{E791-sigma,FOCUS}, where it is assumed the three final particles interact only in pairs with the third as a spectator. If three body forces are needed, they are usually assumed to have constant matrix elements and so populate the Dalitz plot according to phase-space. In the analysis of the Fermilab E791 results~\cite{E791-sigma} with $\sim1100$ $D^+\to\pi^+\pi^+\pi^-$ events, the known resonances that couple to $\pi\pi$ ---
the $\rho$, $f_2(1270)$, etc. --- are included. The resulting fit is poor at low $\pi\pi$ masses. However, this is dramatically improved if an $I=J=0$ resonance of mass $(478 \pm 24 \pm 17)$ MeV and width $(324 \pm 42 \pm 21)$ MeV is added in Breit-Wigner form. Hence the E791 group claim to have confirmed the $\sigma$ resonance~\cite{E791-sigma}. This is to forget that, assuming an isobar model, one has by definition now determined the $I=J=0$ $\pi\pi\to\pi\pi$ interaction. Though the phase-shift is large, it is undoubtedly not given alone by a simple Breit-Wigner form with the claimed mass and width.

With $\sim1500$ events on this same decay, the FOCUS group~\cite{FOCUS} at Fermilab has confirmed the E791 data and analysis. But if instead the low mass $S$-wave $\pi\pi$ interaction is parametrised using the Anisovich and Sarantsev description, then the fit to the $D\to 3\pi$ Dalitz plot is even better. 
Though here and in Ref.~27, we have queried the wide $f_0(980)$ of AS, this matters little for $D\to 3\pi$ decay. Their amplitudes and those of ReVAMP are virtually the same from $\pi\pi$ threshold to 900 MeV (see Fig.~1 of Ref. 27). What 
Malvezzi and collaborators~\cite{FOCUS} have shown is that the $\pi\pi$ final state interactions in $D\to 3\pi$ decay are completely consistent with what we know of such interactions from all other processes. They embody nothing new!

Whether there is a $\sigma$ or not as a short-lived resonance is then not a question of whether it is just seen in $D\to 3\pi$ decay, but whether it is there in all other $I=J=0$ $\pi\pi$ final states too. The difficulty in answering the question of whether data on the real axis have sufficient precision to determine the existence of such  a very distant pole~\cite{mrp-whs99} is amply illustrated by the difference between the AS~\cite{AS} and ReVAMP~\cite{revamp} amplitudes. AS have no low mass pole, while ReVAMP does : yet each describes essentially the same experimental results below 900 MeV.  

What we learn is that analysing data on a single channel, where final state interactions are important, cannot be meaningfully done in isolation. Unitarity requires consistency between reactions. Only by analysing data from different processes with the same final states simultaneously can we hope to be able to draw definitive conclusions about the fascinating scalar sector. Since these states
with zero quantum numbers  reflect the nature of the QCD vacuum, further 
study is essential. Precision results from Hall~D at Jefferson Lab should contribute enormously to this endeavour. 

\vspace{5mm}
\section*{Acknowledgments}
I am very grateful to Alex Dzierba for inviting me to give this talk
and to Robert Edwards and Wally Melnitchouk for hosting this most rewarding
meeting on gluonic excitations. It is a pleasure to thank Elena Boglione and
Sandra Malvezzi for exacting discussions on analysing data.
I acknowledge the partial support of the EU-RTN Programme, 
Contract No. HPRN-CT-2002-00311, \lq\lq EURIDICE''.

\end{document}